\documentclass[journal=jacsat,manuscript=article]{achemso}

\usepackage[version=3]{mhchem} 
\usepackage[T1]{fontenc}       

\author{Fabian O. von Rohr}
\affiliation{Department of Chemistry, Princeton University, Princeton, New Jersey 08544, USA}

\author{Huiwen Ji}
\affiliation{Department of Chemistry, Princeton University, Princeton, New Jersey 08544, USA}

\author{F. Alexandre Cevallos}
\affiliation{Department of Chemistry, Princeton University, Princeton, New Jersey 08544, USA}

\author{Tong Gao}
\affiliation{Department of Physics, Princeton University, Princeton, New Jersey 08544, USA}

\author{N. Phuan Ong}
\affiliation{Department of Physics, Princeton University, Princeton, New Jersey 08544, USA}

\author{Robert J. Cava}
\affiliation{Department of Chemistry, Princeton University, Princeton, New Jersey 08544, USA}

\title{High-Pressure Synthesis and Characterization of $\beta$-GeSe - A Semiconductor with Six-Rings in an Uncommon Boat Conformation}
\keywords{black phosphorus, germanium selenide, semiconductor}

\begin{document}




\begin{abstract}
Two-dimensional materials have significant potential for the development of new devices. Here we report the electronic and structural properties of $\beta$-GeSe, a previously unreported polymorph of GeSe, with a unique crystal structure that displays strong two-dimensional structural features. $\beta$-GeSe is made at high pressure and temperature and is stable under ambient conditions. We compare it to its structural and electronic relatives $\alpha$-GeSe and black phosphorus. The $\beta$ form of GeSe displays a boat conformation for its Ge-Se six-ring, while the previously known $\alpha$ form, and black phosphorus, display the more common chair conformation for their six-rings. Electronic structure calculations indicate that $\beta$-GeSe is a semiconductor, with an approximate bulk band gap of $\Delta~\approx$~0.5 eV, and, in its monolayer form, $\Delta~\approx$~0.9 eV. These values fall between those of $\alpha$-GeSe and black phosphorus, making $\beta$-GeSe a promising candidate for future applications. The resistivity of our $\beta$-GeSe crystals measured in-plane is on the order of $\rho \approx$ 1 $\Omega$cm, while being essentially temperature independent. 
\end{abstract}

\section{Introduction}
Two-dimensional materials such as graphene and transition metal dichalcogenides have shown good potential as the basis for future electronics technologies \cite{Liu15,Ling15}. Similarly, black phosphorus (black-P) is considered to be a promising candidate for many applications in electronics \cite{electronics} and optoelectronics \cite{optoelectronics,optoelectronics2,optoelectronics3}, because, for a small band-gap semiconductor, it displays high mobility charge carriers and can easily be exfoliated \cite{highmob,highmob2,highmob3,exfoliat}. Similar to graphite, due to the weak interlayer van der Waals interactions, black phosphorus can be reduced to a single atomic bi-layer (here we refer to this as a "monolayer"), commonly referred to as phosphorene \cite{electronics}. The physical properties of phosphorene differ significantly from its bulk counterpart \cite{Guo15}. The intrinsic lack of a band gap in graphene has been an obstacle for the construction of devices, inspiring work to overcome this deficit through doping or strain (see, e.g., references \citenum{graphene,graphene2,graphene3,graphene4}). Black phosphorus on the other hand has a small intrinsic band-gap, which can be tuned into a Dirac semimetal \cite{semimetal}.\\ \\
Black phosphorus is the thermodynamically stable phosphorus form, but it oxidizes when exposed to water and oxygen \cite{stability,stability2}. Therefore, related and more stable layered semiconductors are of significant interest. Similar to the relationship between group IV Si and III-V semiconductors like GaAs, binary IV-VI compounds such as GeS, GeSe, SnS and SnSe are isoelectronic and isostructural to group V black phosphorus \cite{Schnering}. These compounds also have the potential for application. For example SnSe has an extraordinarily high thermoelectric figure of merit along its $b$-axis \cite{SnSe}, and monolayer and double-layer SnSe and GeSe have been considered as promising materials for ultrathin-film photovoltaic applications \cite{Schaak,Tilley2009,Eychmu2008,Brutchey2010}. Here we report the unique structure and basic electronic properties of a previously unreported layered form of GeSe, which we designate as $\beta$-GeSe. It is stable under ambient conditions, although we synthesized it at high-pressure and high-temperature. We further show how black phosphorus, $\alpha$-GeSe, and $\beta$-GeSe are structurally and electronically related, and that the IV-VI-semiconductors can be considered as pseudo group-V-elements.
\section{Experimental}
\textbf{Synthesis.} Polycrystalline $\alpha$-GeSe was synthesized by heating Germanium pieces (purity 99.999\%) and polished selenium shot (purity 99.95\%) in an evacuated quartz tube to 750 $^\circ$C where it was annealed for 48h; the product was cooled to room temperature at 180 $^\circ$C/h. Single crystals of $\alpha$-GeSe were obtained by the vertical Bridgman method. The polycrystalline $\alpha$-GeSe was ground to a fine powder and filled in a carbon-coated quartz glass tube with a diameter of 5 mm. The sample was heated to 750 $^\circ$C and moved through the furnace at 0.2 mm/h. Polycrystalline samples, single crystals of $\beta$-GeSe, and single crystals of black phosphorus were obtained by high-pressure, high-temperature synthesis. Fine powder of polycrystalline $\alpha$-GeSe or of dry red phosphorus (purity 99.99\%) were placed in a boron-nitride crucible of 5 mm diameter and 8 mm length. For $\beta$-GeSe, a pressure of 6 GPa, and for black phosphorus a pressure of 1.5 GPa\cite{black-P} in a cubic multi-anvil module (Rockland Research Corp.) and a temperature of 1200 $^\circ$C was applied for up to one hour. Polycrystalline samples were obtained when the product was quenched. Large single crystals for both $\beta$-GeSe and black phosphorus were obtained when the material was cooled at 50 $^\circ$C per minute. \\ \\
\textbf{Characterization.} Single crystal X-ray diffraction was performed using a Bruker D8 VENTURE diffractometer equipped with a Photon 100 CMOS detector, with Mo K$_\alpha$ and a graphite monochromator ($\lambda$ = 0.71073 \AA) at room temperature.  No diffuse scattering was observed. The unit cell determination and subsequent data collection, integration, and refinement were performed with the Bruker APEX II software package. The crystal structure was determined using SHELXL-2013 software implemented through the WinGX Suite. Powder x-ray diffraction patterns were obtained in a Bragg-Bretano reflection geometry on a Bruker D8 Advance Eco with Cu $K_\alpha$ radiation and a LynxEye-XE detector. Refinements were performed using the FullProf program suite \cite{Fullprof}.
The temperature-dependent resistivity and magnetoresistance (MR) measurements were performed using the resistivity option in a \textit{Quantum Design} Physical Property Measurement System (PPMS). For most resistivity measurements, a standard 4-probe technique was employed using 25 $\mu$m diameter gold wire bonded with silver paint. The surface of $\beta$-GeSe crystals was found to not bond well with silver paint and silver epoxy. Therefore, indium bonding was employed instead. The applied current varied with the sample resistance - for resistances below 1 kOhm, we used 100 $\mu$A, and for resistances higher than 1 kOhm, the current was kept below 10 $\mu$A. 
\\ \\
\begin{figure}
	\centering
	\includegraphics[width=0.5\linewidth]{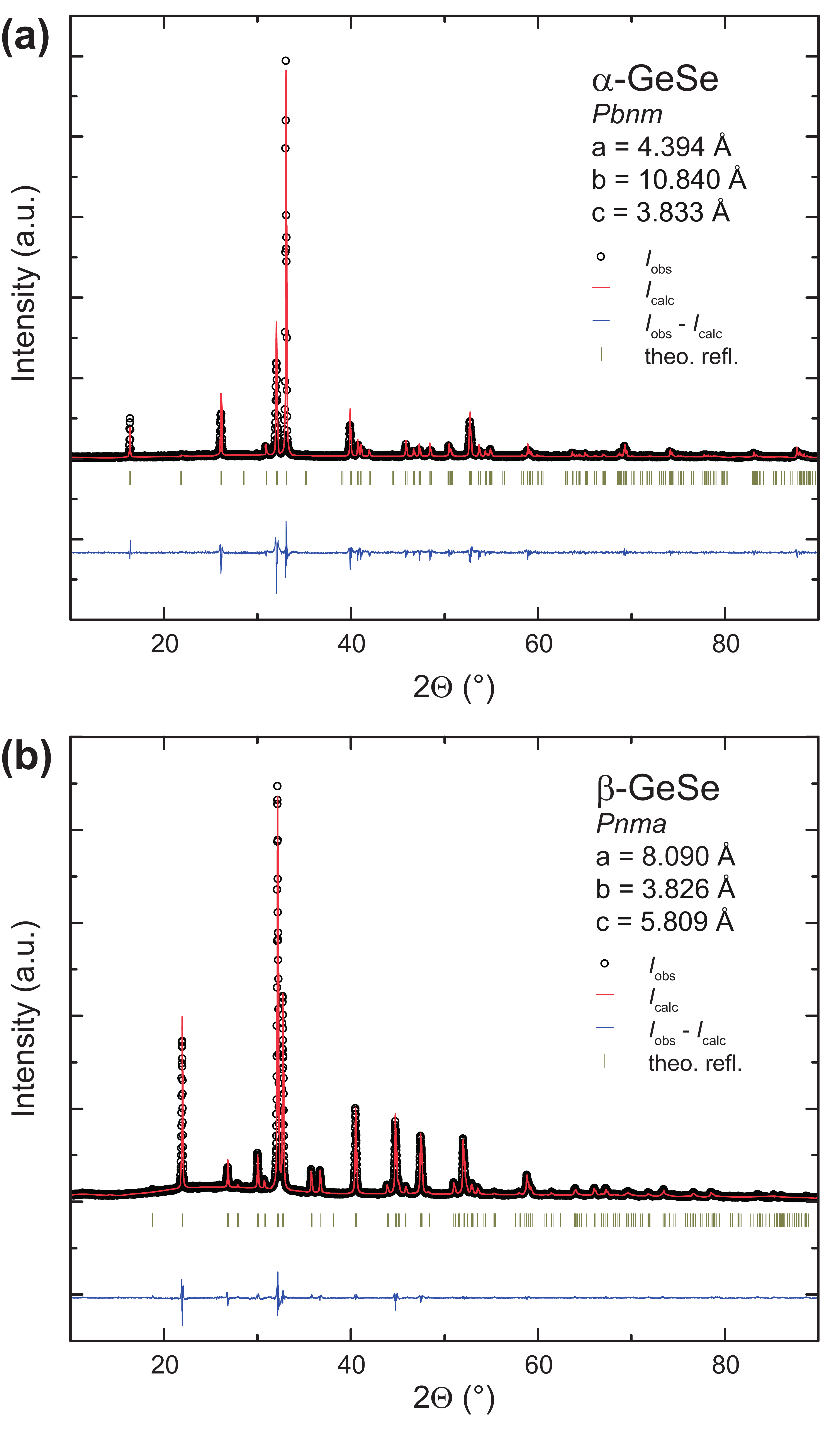}
	\caption{The powder X-ray diffraction (PXRD) patterns of (a) $\alpha$-GeSe and (b) $\beta$-GeSe at ambient temperature and pressure. The black circles represent the observed intensity ($I_{\rm obs}$), the red line shows the Le Bail fits according to the structural models ($I_{\rm calc}$), and the green tics represent the positions of the Bragg reflections.}
	\label{fig:1}
\end{figure}  
\textbf{Calculations.} The electronic band structure calculations were performed using density functional theory (DFT), via the WIEN2K code with a full-potential linearized augmented plane wave and local orbitals basis \cite{Wien2K,Wien2K2,Wien2K3}, together with the Perdew-Burke-Ernzerhof parametrization of the generalized gradient approximation. The mBJ functional was used due to its improved accuracy for band gaps. The plane-wave cutoff parameter R$_{\rm MT}$K$_{\rm max}$ was set to 7, and the reducible Brillouin zone was sampled by 2000 k points; spin-orbit coupling was included. Monolayer calculations were performed using an artificially constructed unit cell, retaining maximal symmetry conservation, with a single Ge-Se layer and 40 \AA \ of vacuum.
\section{Results and Discussion}
\begin{figure} [h!]
	\centering
	\includegraphics[width=0.6\linewidth]{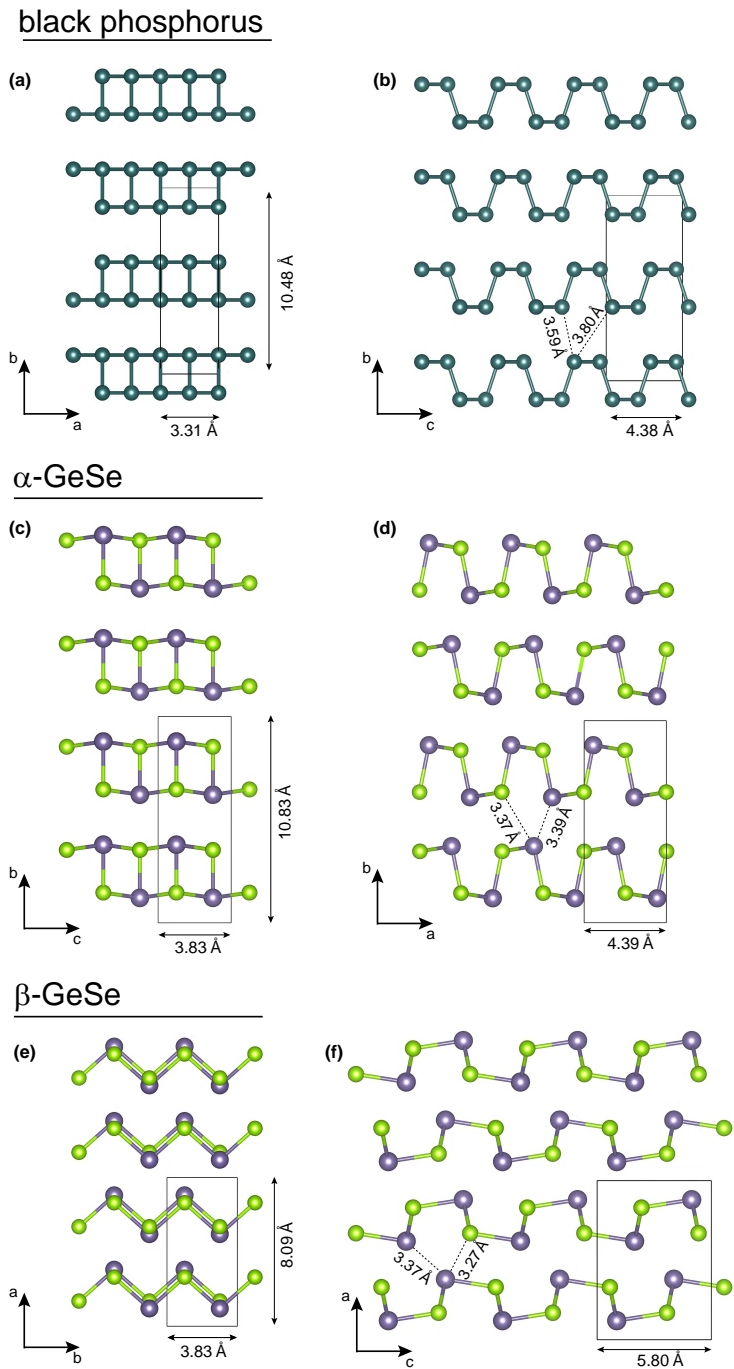}
	\caption{Comparison of the crystal structures of (a,b) black phosphorus, (c,d) $\alpha$-GeSe, and (e,f) the new $\beta$-GeSe polymorph. Phosphorous atoms dark green, germanium atoms grey, selenium atoms bright green.}
	\label{fig:2}
\end{figure}
\begin{figure}
	\centering
	\includegraphics[width=0.6\linewidth]{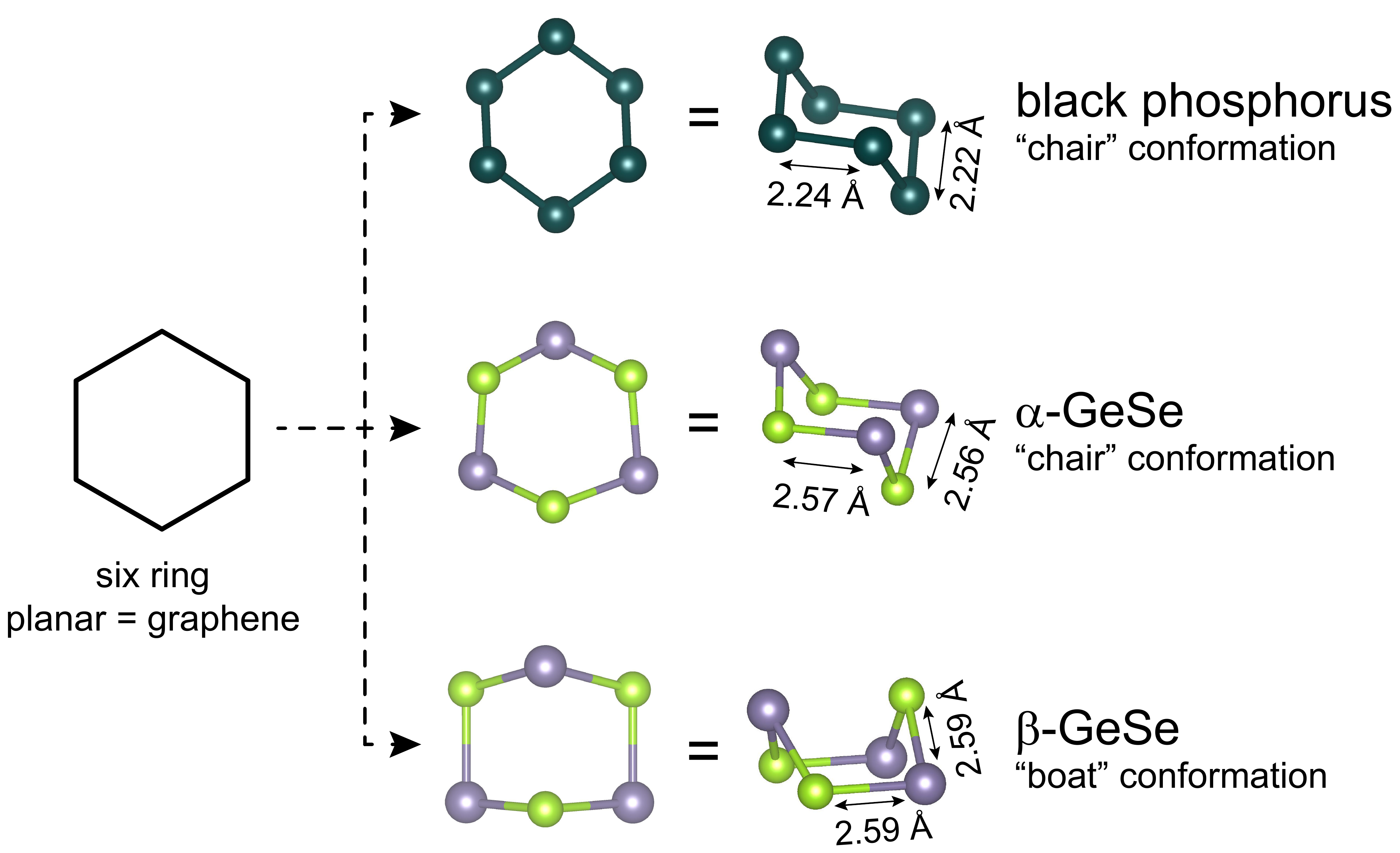}
	\caption{Building blocks of graphene, black phosphorus, $\alpha$-GeSe, and $\beta$-GeSe.}
	\label{fig:3}
\end{figure}
\textbf{Structure and Chemical Bonding.} Both polymorphs, $\alpha$-GeSe and $\beta$-GeSe, were obtained as silvery crystals. They are both strongly layered with well-defined cleavage planes and large shiny surfaces. The crystal structure of the $\beta$-GeSe polymorph was determined using single crystal X-ray diffraction data. $\beta$-GeSe crystallizes in the orthorhombic space group \textit{Pnma}. The crystallographic details are summarized in table \ref{tab:1}, and the refined final structural parameters are listed in table \ref{tab:2}. The phase purity of the crystals used in the characterization studies was confirmed by powder X-ray diffraction (PXRD) on ground crystals. The PXRD patterns for $\alpha$-GeSe and $\beta$-GeSe are shown in figure \ref{fig:1}; all peaks can be indexed according to their respective structural models, (strong preferred orientation of both powders is present) and no impurity phases are observed. The cell parameters for $\alpha$-GeSe are modeled after reference \citenum{Schnering}; this material was found to lose its crystallinity upon grinding, presumably due to the introduction of stacking faults in the layered structure; the $\beta$-GeSe polymorph appears to be more resistant to such faulting. We note that figure \ref{fig:1} is also a "before and after" representation of the high-pressure synthesis, because $\alpha$-GeSe was used as the precursor to synthesize $\beta$-GeSe.\\ \\ 
In figure \ref{fig:2}, we compare the crystal structures of black phosphorus, $\alpha$-GeSe, and $\beta$-GeSe along the main crystallographic directions. $\alpha$-GeSe is isostructural and isoelectronic to black phosphorus. Its structure consists of layers of buckled six-rings in a so-called chair conformation, containing alternating germanium and selenium atoms on the vertices. The structures of black phosphorus and $\alpha$-GeSe can be considered as distorted versions of the rock-salt structure. By comparing the c/a ratios of all column V and IV-VI compounds with this structure, it can be seen that the structures become increasingly more rocksalt-structure-type-like in the sequence: black phosphorus, GeS, $\alpha$-GeSe, SnS, and SnSe \cite{Schnering}. \\ \\
\begin{table} [H]
	\begin{center}
		\begin{tabular}{| c | c |}
			\hline
			\ \ Parameters \ \ & \ \ $\beta$-GeSe \ \ \\
			\hline \hline
			Crystal system & orthorhombic \\
			space group & \textit{Pnma} (No. 62) \\
			lattice parameters [\AA] & $a$ = 8.0892(6) \\
			& $b$ = 3.8261(3) \\
			& $c$ = 5.8089(5) \\
			cell volume [\AA]$^3$ & 179.79(2)\\
			formula units/cell & 4 \\
			$\rho_{\rm cal}$ [g \ cm$^{-3}$] & 5.599  \\
			$\mu$ [mm$^{-1}$] & 36.688 \\
			crystal size [mm] & 0.077 x 0.028 x 0.012\\ 
			F(000) & 264.0\\
			radiation type & Mo K$_{\alpha 1}$ $\lambda$ = 0.71073 \\
			$\Theta$ range [$^{\circ}$] & 4.319 - 36.411 \\
			index ranges & -13 $\le$ h $\le$ 13 \\
			& -6 $\le$ k $\le$ 6 \\
			& -9 $\le$ l $\le$ 8 \\
			observed reflections &  5634\\
			independent reflections ($\le$ 2$\sigma$) & 408 \\
			R$_{\rm int}$ & 0.0555\\
			R$_{\rm \sigma}$ & 0.0282\\
			refined parameters & 13\\
			GOF & 1.133\\
			\textit{R}1 (all data) & 0.0452\\
			\textit{R}1 ($\ge$ 4$\sigma$) & 0.0310\\
			\textit{wR}2 (all data) & 0.0498\\
			\textit{wR}2 ($\ge$ 4$\sigma$) & 0.0527\\
			max/min residual electron density [e \AA$^{−3}$] & 1.880/-1.171\\
			\hline
		\end{tabular}
		\caption{Details of the single-crystal X-ray refinement for $\beta$-GeSe.}
		\label{tab:1}
	\end{center}
\end{table}

\begin{figure}
	\centering
	\includegraphics[width=0.7\linewidth]{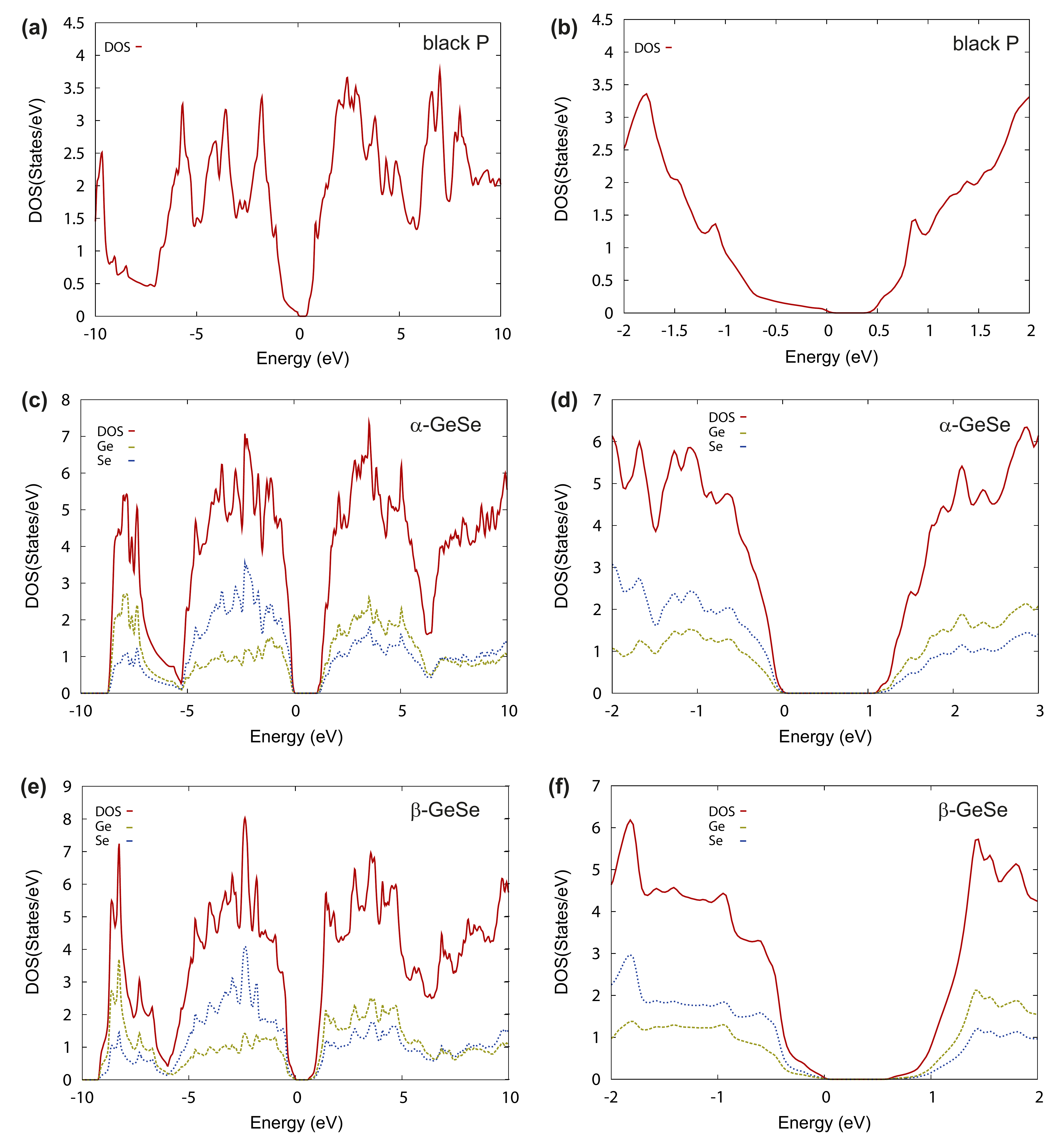}
	\caption{Calculated density of states from -10 eV to +10 eV and in the nearer vicinity around the Fermi-level of (a,b) black phosphorus, (c,d) $\alpha$-GeSe, and (e,g) $\beta$-GeSe.}
	\label{fig:4}
\end{figure}

\begin{table} [h!]
	\begin{center}
		\begin{tabular}{| c || c | c | c | c | c |}
			\hline
			atom & Wyckoff symbol & x & y & z & U$_{\rm eq}$ [\AA$^2$] \\
			\hline \hline
			Ge & 4c & 0.11966(5) & 0.2500 & 0.82830(9) & 0.01383(12) \\
			Se & 4c &  0.16946(5) &  0.2500 &  0.38830(6) & 0.01091(11) \\
			
			\hline
		\end{tabular}
		\caption{Positional coordinates of the atoms in $\beta$-GeSe, in space group \textit{Pnma} at ambient temperature and pressure, and their isotropic thermal displacement parameters.~The unit cell dimensions are [\AA] $a$ = 8.0892(6), $b$ = 3.8261(3), and $c$ = 5.8089(5).} 
		\label{tab:2}
	\end{center}
\end{table}
The $\beta$-GeSe polymorph crystallizes in a distinct crystal structure that is highly layered, but can be considered as more three-dimensional than $\alpha$-GeSe. $\alpha$-GeSe undergoes a transition to the rocksalt structure at high temperatures \cite{Wiedemeier75} and high pressure \cite{rock_GeSe}, but the high-temperature and intermediate pressure polymorph $\beta$-GeSe displays fewer features of the rocksalt structure than does $\alpha$-GeSe. While the building blocks of $\alpha$-GeSe and black phosphorus are six-rings with the vertices in a chair conformation, the structure of $\beta$-GeSe is made up of six-rings with the vertices in a boat conformation. For organic six-ring compounds such as cyclohexane, the chair conformation is commonly found to be the most stable, while the boat conformation is higher in energy. The boat conformation in $\beta$-GeSe appears to be stabilized through a slight decrease of the distance between the layers - the Ge-Se distance across the layers is $d$(Ge-Se) $\approx$ 3.27 \AA \ for $\beta$-GeSe while it is $d$(Ge-Se) $\approx$ 3.37~\AA \ for $\alpha$-GeSe (compare reference \citenum{Schnering}). This causes furthermore a slight shift in the six-ring, resulting in a slightly asymmetrical boat conformation. These building blocks are illustrated in figure \ref{fig:3}. Six-rings that are in a boat conformation are not known for any structure of phosphorus or its other binary IV-VI equivalents. \\ \\
\textbf{Electronic structures.} In figure \ref{fig:4}, we show the calculated electronic density of states within 10 eV of the Fermi level (E$_{\rm F}$), and in the nearer vicinity for black phosphorus (a,b), $\alpha$-GeSe (c,d) and $\beta$-GeSe (e,f). The contributions of germanium and selenium are of very similar magnitude and distribution around E$_{\rm F}$ in both forms of GeSe, an indication of very covalent bonding - no obvious charge transfer can be observed. This strong covalency and lack of charge transfer suggests that both $\alpha$-GeSe and $\beta$-GeSe can be considered as pseudo group-V-elements, making them electronic analogs to black phosphorus. The overall density of states is found to change only slightly when going from $\alpha$-GeSe to $\beta$-GeSe. The $\beta$-GeSe polymorph has a smaller calculated band gap, which may be due to its slightly more three-dimensional structure, which leads to a better overlap of the orbitals between the layers. \\ \\
\begin{figure}
	\centering
	\includegraphics[width=0.7\linewidth]{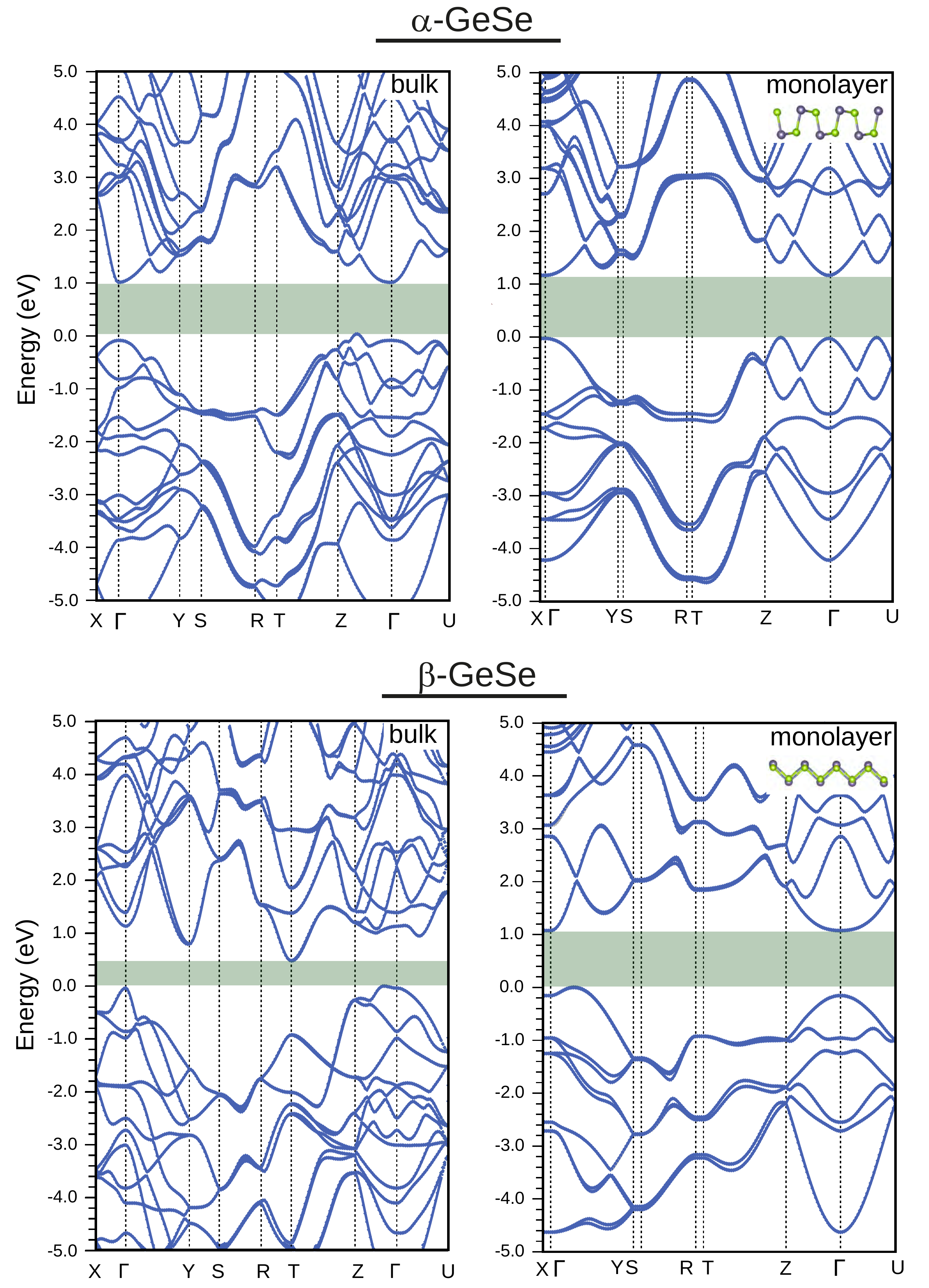}
	\caption{Calculated band structures of bulk and monolayer $\alpha$-GeSe and $\beta$-GeSe. The band gaps are emphasized with green shaded areas.}
	\label{fig:5}
\end{figure}
The results of the band structure calculations for $\alpha$-GeSe and $\beta$-GeSe are shown in figure \ref{fig:5}. The high-symmetry paths in the Brillouin zone were chosen for two structurally similar sequences in order to allow for comparison. The calculated electronic band gap for bulk $\beta$-GeSe ($\Delta \approx$ 0.5 eV) is smaller than that for $\alpha$-GeSe ($\Delta \approx$ 0.8 eV). Both gaps are larger than the calculated band gap for black phosphorus, found by our methods to be $\Delta \approx$ 0.3 eV for the bulk material (compare, e.g. references \citenum{BP-gap,BP-gap2}). The band gap energies range from the infrared to the visible; both $\alpha$-GeSe and $\beta$-GeSe are calculated to have indirect band gaps. We find the calculated band gaps for both $\alpha$-GeSe and $\beta$-GeSe monolayers to be larger than those of the bulk materials but approximately equal to each other, in the 0.9 to 1.1 eV range. The calculated gap for monolayer $\alpha$-GeSe is direct, while that for $\beta$-GeSe is calculated to be indirect. We note, however that the calculated direct band gap at the $\Gamma$-Point for $\beta$-GeSe is only slightly larger in energy than the indirect gap and thus experiments will be needed to resolve the actual electronic structure. For comparison with other members of this structural family, see reference \citenum{Gomes15}. \\ \\
\begin{figure} [H]
	\centering
	\includegraphics[width=0.5\linewidth]{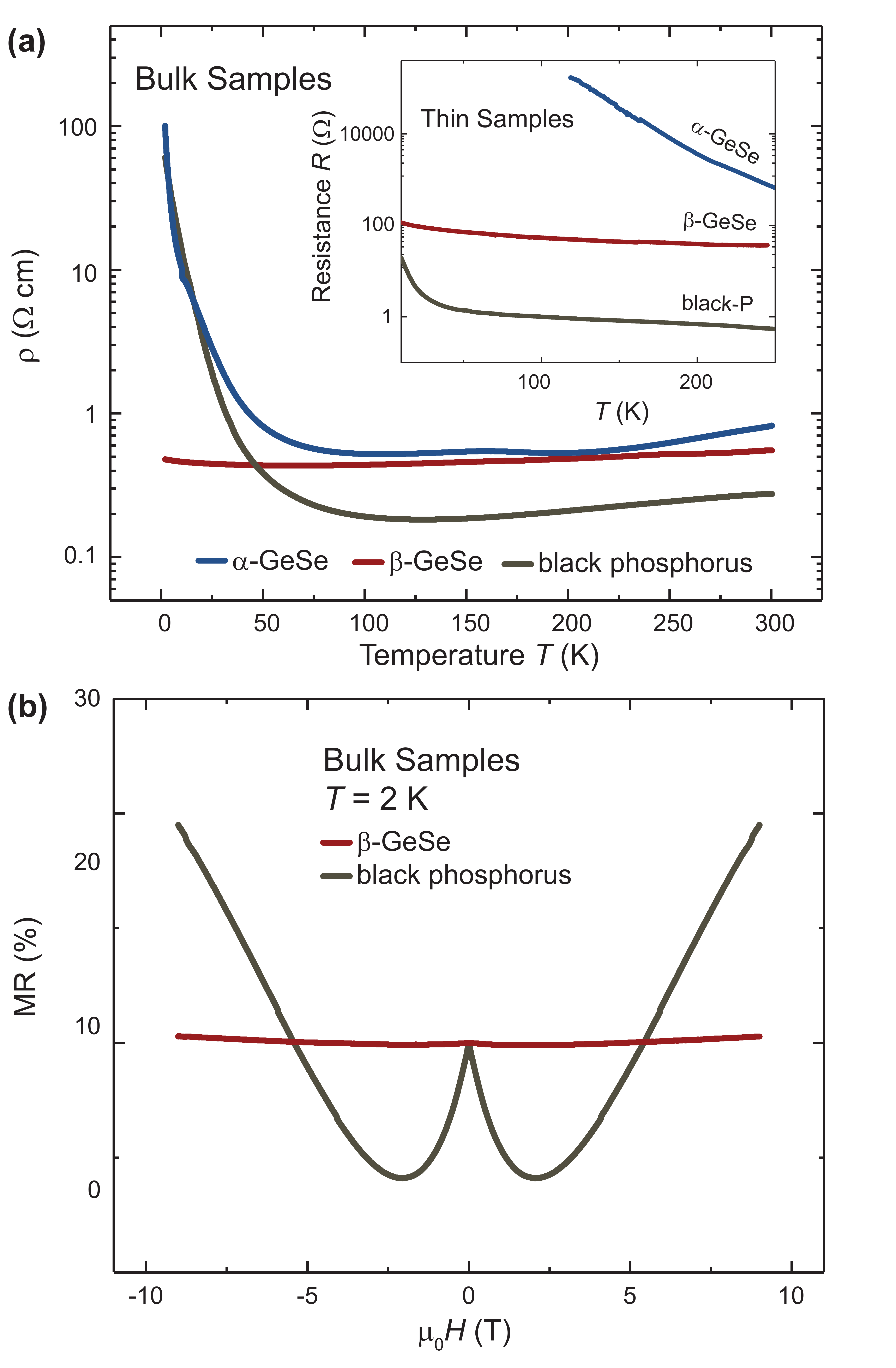}
	\caption{Electronic transport properties of black phosphorus, $\alpha$-GeSe, and $\beta$-GeSe for bulk materials and thin samples (insert). (b) Magnetoresistance of $\beta$-GeSe and black phosphorus at $T =$ 2 K.}
	\label{fig:6}
\end{figure}
\textbf{Transport properties.} In figure \ref{fig:6}(a), we show the temperature dependent resistivities for black phosphorus, $\alpha$-GeSe, and $\beta$-GeSe between $T =$ 300 K and 2 K. The observed resistivities at room temperature are $\rm \rho_{black-P}(300 K) \approx$ 0.27 $\Omega \ {\rm cm}$ for black phosphorus, $\rm \rho_{\alpha-GeSe}(300 K) \approx$ 0.82 $\Omega \ {\rm cm}$ for $\alpha$-GeSe, and  $\rm \rho_{\beta-GeSe}(300 K) \approx$ 0.55 $\Omega \ {\rm cm}$ for $\beta$-GeSe. The resistivity of black phosphorus and its temperature-dependence is in good agreement with earlier reports \cite{BP-gap}. The resistivity of $\beta$-GeSe falls between those of $\alpha$-GeSe and black phosphorus. Black phosphorus and $\alpha$-GeSe have resistivites that increase with decreasing temperature at low temperatures while the resistivity of $\beta$-GeSe is almost temperature independent down to low temperatures. The behavior observed for $\alpha$-GeSe and black phosphorus above 200 K and is characteristic of heavily doped semiconductors: as the product of the carrier density and the mobility of the carriers, and the slight decrease of the resistivities with temperature in these two materials may be attributed to high mobility of the carriers - a possibility that has been discussed in detail for black phosphorus (see, e.g., reference \citenum{highmob}). The observed relatively high but temperature independent resistivity of $\beta$-GeSe, compared to its calculated electronic structure showing that it should be a 0.5 eV semiconductor, suggests that the observed resistivity is dominated by defect-state conduction. Future work will be required to experimentally confirm the calculated electronic structure and optimize the resistivity of this compound for future applications. \\ \\
The temperature dependent resistances of relatively thin crystals of black phosphorus, $\alpha$-GeSe, and $\beta$-GeSe are shown in the inset of figure \ref{fig:6}(a); the approximate crystal thicknesses are 0.1 mm (100 microns). Similar to the bulk materials, the resistance of $\beta$-GeSe falls between those of $\alpha$-GeSe and black phosphorus. We find that $\alpha$-GeSe becomes strongly insulating when the size of the crystal is heavily reduced, reaching values of over 10,000 $\Omega$ below $T =$ 150 K. Black phosphorus and $\beta$-GeSe, however, retain measurable resistances even for relatively thin samples, an indication of their potential to serve in advanced electronic devices.\\ \\
In figure \ref{fig:6}(a), we show the magnetoresistance (MR) of $\beta$-GeSe and black phosphorus at $T =$ 2 K. (The MR of $\alpha$-GeSe could not be measured by our methods due to the large resistance of the sample.) Black phosphorus shows a non-linear increasing MR with field, reaching a value of $\approx$10\% at $\mu_0 H =$ 9 T, with a parabolic behavior in the high-field region indicative of conduction via multiple types of carriers. The MR of $\beta$-GeSe, on the other hand is linear, very small, and almost field independent. This is consistent with our speculation about defect state-dominated conduction in these first samples of $\beta$-GeSe, but further, more detailed study will be required to interpret the electronic character of this compound. 
\section{Summary and Conclusion} 
We have described the synthesis and basic electronic characterization of a previously unreported form of GeSe, which we designate as $\beta$-GeSe. The crystal structure is orthorhombic, and strongly layered, with Ge-Se distances within the layers of $d$(Ge-Se) $\approx$ 2.59 \AA \ and between the layers of $d$(Ge-Se) $\approx$ 3.27 \AA. The layers of $\beta$-GeSe consist of six-rings with the vertices arranged in a boat conformation. This is in contrast to $\alpha$-GeSe and black phosphorus where the atoms in the layers arrange in a six-ring with the vertices in a chair conformation. The observed boat conformation for $\beta$-GeSe is less frequently observed for isolated six-ring structures, and we propose that it is stabilized by the interlayer interactions. Electronic structure calculations show that both $\alpha$-GeSe and $\beta$-GeSe can be considered as pseudo group-V-elements due to the mostly covalent character of the bonding at the Fermi-level. Our calculations and experiments suggest that $\beta$-GeSe is highly analogous to black phosphorus and $\alpha$-GeSe and falls electronically between them in both bulk and monolayer form. This makes $\beta$-GeSe a promising candidate for future device applications. 
\section*{Acknowledgments}
This work was supported by the Gordon and Betty Moore Foundation, EPiQS initiative, Grants GBMF-4412 and GBMF-4539. The authors thank Lukas M\"uchler for helpful discussions.

\end{document}